\documentclass[
    ,final            
  ]
  {aipproc}

\layoutstyle{8x11double}

\begin{document}

\title{Anharmonic effects on infrared spectra of GaAs and GaP}

\author{Hadley M. Lawler}{
  address={Department of Physics, University of Maryland, College Park, MD 20742}
}

\author{Eric L. Shirley}{
  address={Optical Technology Division, NIST, Gaithersburg, MD 20899-8441}
}

\begin{abstract}
With a study of bulk GaAs and GaP, we report detailed infrared spectra and their temperature dependence from first principles.  Fine features and low-temperature trends in the predicted response are confirmed by experiment.  The spectra are calculated from the far-infrared through twice the frequency of the zone-center optical phonon.  The treatment relies on recent developments in the theories of dielectric polarization and the phonon-phonon interaction.
 \end{abstract}

\maketitle


  
\indent  {\it{Ab initio}} prediction of the optical response of simple solids over a very broad spectral range is a significant achievement of theory and is founded on a quantitative understanding of the electronic structure~\cite{genspec}.  In a related advance, complete characterizations of lattice dynamics in the harmonic approximation, where phonons do not interact, are now routine~\cite{phonrev}.

\indent  The harmonic approximation, however, is insufficient for a close inspection of the frequency dependence of the dielectric response in the infrared (IR).  For instance, in the absence of direct multiphonon transitions, the harmonic approximation predicts infinitely sharp absorption at $\omega_{TO}$, the frequency of the long-wavelength, transverse optical phonon, or the dispersion oscillator, and nowhere else~\cite{Born,Giannozzi}.  Measured spectra, however, show finite linewidth and asymmetric lineshape of the IR-active phonon and very broad background absorption, which can exhibit fine features and singularities~\cite{Thomas1,PalikP,PalikAs}, and temperature-dependent absorption in the far-infrared~\cite{Stolen}.  These effects can be explained by incorporating the phonon-phonon interaction into the theory of the optical response, where anharmonic terms in a crystal's interionic potential cause the phonons to interact~\cite{Vin,Mar2,Cowl2}.  As an example of this, an infrared photon can, through anharmonic corrections to the IR-active phonon's quantum state, be absorbed by a pair of phonons whose momenta sum to zero and whose energies sum to that of the photon.  These sum processes endow the spectra with rich structure.  At nonzero temperatures, the onset of difference processes, where one phonon is emitted and another is absorbed along with the photon, give a strong temperature dependence to the far-infrared absorption.     

\indent  Realistic, observable quantities arising from phonon-phonon interactions are now calculated as well as measured, such as  the linewidth, asymmetric shape and temperature dependence of Raman modes~\cite{Deb1,Lang,Deb2,GaP,GaN,AlAs}.  In this Letter, the treatment of the phonon-phonon interactions, or anharmonic effects, closely follows the theoretical studies of Raman spectra, where knowledge of a phonon self-energy is of central importance.  Examination of the phonon self-energy's effects in the dielectric response is interesting in its own right, and the ability to calculate it may aid in the design of infrared optical components.  Phonon-induced dipoles are given by the Born effective charges. These charges enhance the longitudinal optical phonon frequency while coupling the transverse optical phonon to infrared photons, and they are calculated with the modern theory of polarization~\cite{Van,Resta,Gonze}.  With these charges and the IR-active phonon's self-energy, the spectral profile and the oscillator strength can be calculated, and all the optical constants can be obtained.  

\indent  Figures~1 and~2 summarize our results and their comparison with measured spectra. Their examination suggests that two-phonon processes are present over the entire infrared spectral range through 2$\omega_{TO}$.  Below $\omega_{TO}$, we predict the measured~\cite{Stolen} low-energy absorption tail well into the far-infrared spectrum, and the expected temperature dependence.  The lineshape of the IR-active phonon is strikingly reproduced.  Between $\omega_{TO}$ and 2$\omega_{TO}$, absorption and transparency features demonstrate a convincing correspondence with measured data.  These are associated, respectively, with critical points and gaps in the zero-momentum, two-phonon density of states.  There is more structure to the two-phonon spectrum of GaP than of GaAs, including narrow bands of suppressed absorption, because its larger ionic mass mismatch widens the energy difference between various phonon branches. The data for GaAs are from Ref.~\cite{PalikAs} and references therein.  Some data points are from fits to an oscillator model, and are ``generally not determined better than ${\pm}~30~\%$''~\cite{PalikAs}.  The reststrahlen and low-frequency data for GaP are from Ref.~\cite{PalikP} and references therein.  The low-temperature and room-temperature data points above the reststrahlen are from Ref.~\cite{Thomas1}, and the data are from transmission measurements on a high-purity sample.

\begin{figure}
  \includegraphics[height=.5\textheight]{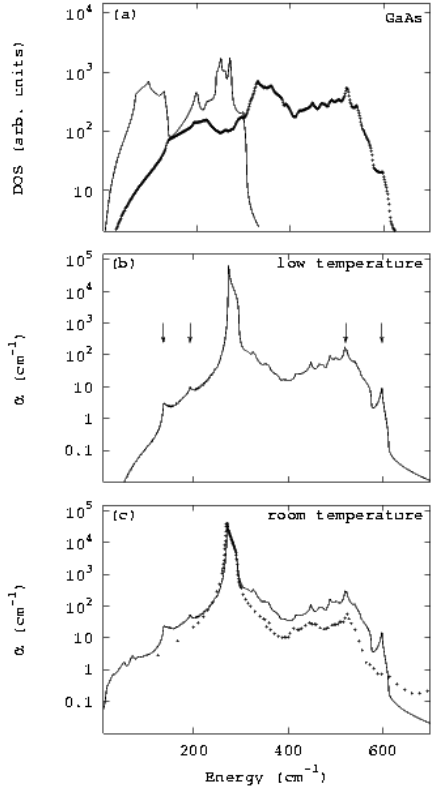}
  \caption{The one-phonon (line) and zero-momentum two-phonon ($+$) densities of states are plotted in (a) for GaAs.  In (b) and (c) the calculated absorption coefficient is plotted at 0~K and 300~K, respectively.  Plotted theoretical spectral features have been moved to 0.98 of their calculated position, in cm$^{-1}$, to facilitate comparison with the experimental 300~K spectra, which are plotted in (c) ($+$).  The arrows index, from left to right, the two-phonon critical points:  $TA({ L})+TA({ L}),~TA({ X})+TA({ X}),~TO({ L}) + TO({ L}),$ and $LO({ {\Gamma}}) + LO({ {\Gamma}})$. The optical response is characterized with the absorption coefficient, $\alpha({\nu})=4{\pi}{\nu}k(\nu)$, where $k(\nu)$ is the index of absorption, and $\nu$ is photon energy in cm$^{-1}$.  The data are from Ref.~\cite{PalikAs}.}
\end{figure}

\begin{figure}
  \includegraphics[height=.5\textheight]{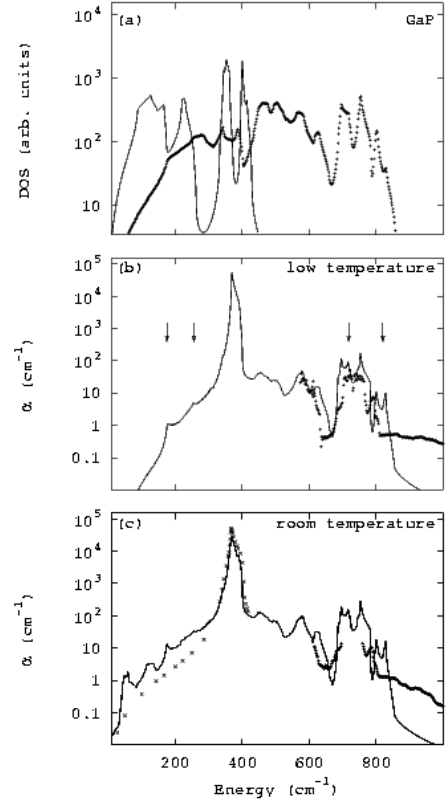}
  \caption{Same as in Fig.~1, for GaP, with the exception of the overtone at ${\bf L}$, as noted in the text.  The data are from Ref.~\cite{Thomas1} (+), and Ref.~\cite{PalikP} (x) as described in the text.}
\end{figure} 

\indent  The extent of similarity between predicted and observed spectra strongly indicates that modern first-principles calculations and perturbative formalism reflect the microscopic processes governing infrared optical response, and may serve to stimulate further work toward detailed characterization of crystalline materials. 

\indent  Below we summarize the role of phonons in the dielectric response, discuss the anharmonic contributions and their many-body treatment, and finally present results and details of our calculations.

\indent  In GaAs and GaP, infrared optical properties are dominated by phonons.  The Hamiltonian density for a polar crystal in the presence of an electric field, {\bf E}, can be written as

\begin{equation}
H=
{\sum}_{\lambda}\frac{1}{2N\Omega}
(\omega^2_{\lambda}q^2_\lambda
+ \dot{q}^2_{\lambda}) -{\bf{E}}\cdot{\bf{P}}^{mac}
+H_{anh}+H_{mp},
\label{Ham}
\end{equation}
where $q_\lambda$ and $\omega_{\lambda}$ are the normal coordinate and frequency, respectively, of a phonon with wave vector and branch indexed by $\lambda$.  The unit cell volume is $\Omega$ and the number of Bravais lattice sites is $N$. The dimensions of the quantity ${\bf{P}}^{mac}$ are dipole moment per unit volume.  Anharmonic potential energy terms are assumed to be small, and expressed with $H_{anh}$.  Direct dipole transitions to multiphonon states are represented with $H_{mp}$.  This coupling has been observed and calculated~\cite{Lax,Strauchtp}, but is not accounted for here.  

\indent The macroscopic polarization in Eq.~(\ref{Ham}) can be written as,
\begin{equation}
P^{mac}_i=
\sum_{\nu=TO}
\sum_{{\tau}j}\frac{1}{\sqrt{m_\tau}}{Z^{\tau}_{ij}}q_{\nu}\epsilon^{\nu}_{{\tau}j}.
\label{pol}
\end{equation} 
In the above, $m_\tau$ is the mass of ion $\tau$, the phonon normal coordinate index, ${\nu}$, runs strictly over the IR-active phonons, and $\epsilon^{\nu}_{{\tau}j}$ is a component of the polarization of such a phonon.  Cartesian directions are indicated with the indices $i$ and $j$.  The charges are expressed as the change in macroscopic polarization with respect to sublattice displacement:
\begin{equation}
{Z^{\tau}_{ij}}=\Omega\frac{{\partial}P^{mac}_i}{{\partial}u_{{\tau}j}}.
\label{charge}
\end{equation} 
Modern theory gives elegant and tractable prescriptions with which to evaluate $Z^{\tau}_{ij}$~\cite{Van,Resta}.

\indent  Formally, the following expressions can be derived with a formalism analogous to that of Kubo~\cite{Kubo}.  A heuristic oscillator model for the response can be written by expressing the effects of $H_{anh}$ as a complex damping term, or self-energy that modifies $\omega_{TO}$: $\Sigma_{TO}(\omega)=\triangle(\omega) -i\gamma(\omega)$. Substituting Eq.~(\ref{pol}) into Eq.~(\ref{Ham}), solving the driven-oscillator equation of motion for an IR-active phonon's normal coordinate, and then taking the sum indicated in Eq.~(\ref{pol}), the polarization is found to be:
\begin{eqnarray}
P^{mac}_i(\omega)  = \sum_{{\tau}{\tau}'i'jj'}
\frac{{Z^{\tau}_{ij}}{Z^{\tau'}_{i'j'}}
        \epsilon^{\nu}_{{\tau}j}\epsilon^{\nu}_{{\tau'}j'}E_{i'}(\omega)}
      {2\sqrt{m_{\tau}m_{\tau'}}{\omega_{TO}}}~~~~~~~~~~~~~~~~~~~~~~\\
\nonumber
\sum_{{\nu=TO}}\left(
\frac{1}{\omega+\omega_{TO}+
\Sigma_{TO}(\omega)} - 
\frac{1}{\omega-\omega_{TO}-
\Sigma_{TO}(\omega)}
\right)
\end{eqnarray}
The dielectric function follows immediately.  
It is written~\cite{Gonze,Maradudin}:
\begin{eqnarray}
\nonumber
\varepsilon_{ii'}({\omega})=
\varepsilon_{{ii'}}^{\infty} +\frac{4\pi}{\Omega}\sum_{{\tau}{\tau}'jj'}\frac{{Z^{\tau}_{ij}}{Z^{\tau'}_{i'j'}}\epsilon^{\nu}_{{\tau}j}\epsilon^{\nu}_{{\tau'}j'}}{2\sqrt{m_{\tau}m_{\tau'}}{\omega_{TO}}}~~~~~~~~~~~~~~~~~~\\
\sum_{{\nu=TO}}\left(
\frac{1}{\omega+\omega_{TO}+
\Sigma_{TO}(\omega)}
-\frac{1}{\omega-\omega_{TO}-
\Sigma_{TO}(\omega)}
\right)
\label{response}
\end{eqnarray}
with $\varepsilon_{{ii'}}^{\infty}$ representing the high-frequency dielectric constant.  When all the IR-active modes are symmetry-related, as is the case in a zincblende system, their frequencies and self-energies are equal, and denoted by the subscript $TO$ above.  

\indent  Low-temperature phonon-phonon interactions have been the subject of thorough field-theoretic analyses~\cite{Vin,Mar2,Cowl2,Hove,Mar3,Kokkaddee,handbuch}. Modern implementations successfully calculate realistic thermodynamic and spectral properties which originate from such interactions~\cite{Deb3,Strauch}, demonstrating the valuable predictive nature of first-principles studies.  Phonon-phonon scattering events are ranked with a dimensionless expansion parameter, for which Van Hove has posited the ratio of zero-point ion delocalization to nearest-neighbor lengths~\cite{Hove}.  The theory establishes lowest-order contributions to the anharmonic self-energy of a long-wave phonon from three distinct multiphonon processes~\cite{Hove,Mar3,Kokkaddee}.  These are shown in Figure~3.   One of these processes, (a), involves one four-phonon vertex (second order in the expansion parameter), and two, (b) and (c), involve three-phonon vertices (each first order in the expansion parameter) and are second-order in $H_{anh}$.  The four-phonon process and one of the three-phonon processes represent simultaneous absorption and reemission of a phonon by the IR-active phonon.  Such interactions are instantaneous, and therefore independent of frequency, contributing exclusively a line shift~\cite{Cowl3}.  Our interest is in spectral structure and multiphonon processes, so we have neglected these frequency-independent terms.  In any case, our calculations assume the theoretically relaxed rather than actual lattice constant, so the positions of $\omega_{TO}$ and related spectral features are affected by other factors as well.

\begin{figure}
  \includegraphics[height=.1\textheight]{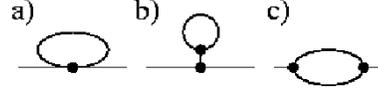}
  \caption{Lowest-order processes contributing to the IR-active phonon self-energy.  The vertex of diagram (a) represents a four-phonon interaction and is related to energy shifts associated with quartic terms in lattice strain.  Diagram (b) expresses shifts associated with thermal expansion.  Only diagram (c) contributes frequency-dependence to the self-energy.}
\end{figure} 

\indent  The remaining term, (c), reflects the IR-active phonon's coupling to phonon-pair states of opposite crystal momenta, ${\bf k}$ and $-{\bf k}$, where their role has been compared to that of a ``viscous medium''~\cite{Born},
and the matrix element between the IR-active phonon and such a state is:
\begin{equation}
M_{\alpha\beta{\bf k}}=\sum_{\tau\tau'ii'}
\frac{{\partial}{D^{\tau\tau'}_{ii'}({\bf k})}}{{\partial}u}
\epsilon^{{\bf k} \alpha}_{{\tau}i}\epsilon^{{-\bf k} \beta}_{{\tau'}i'}.
\end{equation}
The derivative of the dynamical matrix, $D^{\tau\tau'}_{ii'}({\bf k})$, is taken with respect to the sublattice relative coordinate.  Phonon branches are denoted $\alpha$ and $\beta$.
 For a zincblende crystal, the relevant anharmonic contribution to the Hamiltonian is:
\begin{equation}
H_{anh}=
\frac{1}{\sqrt{2\mu}N\Omega}
\sum_{\nu=TO}
\sum_{{\bf k}\alpha\beta}
{M_{\alpha\beta{\bf k}}}
q_{{\bf k}\alpha}q_{{-\bf k}\beta}q_{\nu}
.
\label{anharmonic}
\end{equation}
The prefactor includes the reduced mass, $\mu=(m_{Ga}^{-1}+m_{As}^{-1})^{-1}$, or $\mu=(m_{Ga}^{-1}+m_{P}^{-1})^{-1}$.  The anharmonic interaction in Eq.~(\ref{anharmonic}) leads to a frequency-dependent shift and damping of the IR-active phonon, expressed as the self-energy:
\begin{eqnarray}
\nonumber
\Sigma_{TO}(\omega)=
-\frac{\hbar}{4\mu{\omega_{TO}}}
\frac{1}{N}
\sum_{{\bf k}\alpha\beta}
\frac{|{M_{\alpha\beta{\bf k}}}|^2}{\omega_{{\bf k}{\alpha}}\omega_{-{\bf k}{\beta}}}~~~~~~~~~~~~~~~~~~~~~~~~(8)~~~~~~~~~~~~~~~~~~~~~~~~~~~~~~~~~~~~~~~~~~~~~~~~~~~~~~~~~~~~~~~~~~~~~~\\
\nonumber
\left[
\left(\frac{1+n_{{\bf k}\alpha}+n_{-{\bf k}\beta}}{{\omega_{{\bf k}{\alpha}}}+\omega_{-{\bf k}{\beta}}-\omega-i\eta}\right)+\right.\left(\frac{1+n_{{\bf k}\alpha}+n_{-{\bf k}\beta}}{{\omega_{{\bf k}{\alpha}}}+\omega_{-{\bf k}{\beta}}+\omega+i\eta}\right)~~~~~~~~~~~~~~~~~~~~~~~~~~~~~~~~~~~~~~~~~~~~~~~~~~~~~~~~~~~~~~~~~~~~\\\nonumber
\left.
+\left(\frac{n_{{\bf k}\beta}-n_{{\bf k}\alpha}}{{\omega_{{\bf k}{\alpha}}}-\omega_{{\bf k}{\beta}}-\omega-i\eta}\right)+
\left(\frac{n_{{\bf k}\alpha}-n_{{\bf k}\beta}}{{\omega_{{\bf k}{\beta}}}-\omega_{{\bf k}{\alpha}}-\omega-i\eta}\right)
\right],~~~~~~~~~~~~~~~~~~~~~~~~~~~~~~~~~~~~~~~~~~~~~~~~~~~~~~~~~~~~~~~~
\label{senergy}
\end{eqnarray}
which is derived in detail in Refs.~\cite{Vin,Mar2,Cowl2}.  The summation includes mode frequencies, $\omega_{{\bf k}{\alpha}}$, and Bose occupation factors, $n_{{\bf k}{\alpha}}$.  We work within a frozen-phonon framework~\cite{YandC,us}, using density-functional theory, pseudopotentials and a plane-wave basis~\cite{HK,Hamman,Vanderbilt,KandB}.  The results for the two distorted crystals are subtracted to evaluate the derivative in the dynamical matrix of Eq.~(\ref{anharmonic}) at 122 points within the Brillioun zone.  Fitting 22 inequivalent real-space interatomic force constants facilitated increasing the Brillouin zone sampling to $\sim 10^6$ points.

\indent  In Figures~1 and~2, along with the theoretical and experimental values for the absorption coefficient, we have also plotted the density of momentum-conserving two-phonon states for comparison with the absorption features.  The dispersion-oscillator-two-phonon quasiparticle model we have adopted is shown to agree remarkably with experimental results near the reststrahlen and qualitatively away from it.  The arrows in the figures represent a few sum-process critical points as identified in the caption.  The wave vectors are indicated with the standard notation for the high-symmetry points of a face-centered cubic reciprocal lattice, and the branches are labeled L(T)A(O) for longitudinal (transverse) and acoustic (optical).  The three most energetic modes are denoted optical, the doubly degenerate modes denoted transverse, and the singlets denoted longitudinal.  The identified two-phonon overtones are identical for each material, with the exception of the optical overtone at ${\bf L}$, where the singlet is more energetic in GaP, and the reverse is true for GaAs.  The doublet overtone is indicated for GaAs, and the singlet for GaP. The longitudinal-transverse band crossing in GaAs is a result of its similar ionic masses, as discussed below.  Features above $2\omega_{LO}$ are evident in the theoretical spectra of GaP because of slight overbending in the calculated phonon dispersion.

\indent Above the reststrahlen, the theory reproduces spectral structure better than it does overall amplitude.  For instance, the calculated absorption for GaAs is typically between one half and one quarter of the measured value between 400~cm$^{-1}$ and 600~cm$^{-1}$.  The relative values for the measured absorption are more reliable than their absolute values~\cite{PalikAs}, and this fact may contribute to the discrepancy.  But it may be that the two-phonon features are  not completely described through the IR-active phonon self energy, and that direct multiphonon transitions should be considered.  As can be seen from Eq.~(\ref{senergy}), our model cannot give absorption above $2\omega_{TO}$, where three-phonon and higher-order processes must be considered~\cite{Thomas}.

\indent The temperature dependence of the absorption below 100~cm$^{-1}$ is ascribable to difference processes, where a phonon is absorbed and one of the same momentum but differing branch is emitted.  Such processes are expressed in the third and fourth terms on the right side of Eq.~(\ref{senergy}), and are only possible at temperatures corresponding to finite phonon occupation numbers.  The absorption grows as the material heats and the occupation numbers increase, and this behavior has been measured in GaAs for three far-infrared frequencies~\cite{Stolen}.

\indent The GaP spectrum possesses a richer two-phonon structure than that of GaAs, including narrow bands of weak absorption, a consequence of its larger ionic mass mismatch, which widens the energy difference between various phonon branches.  In the case of similar masses, such as those in GaAs, the optical and acoustic branches approach near degeneracy at high-symmetry wave vectors, thus narrowing the gap between the bands associated with the branches.  The same is true for the transverse and longitudinal optical branches, which, in the case of GaAs, show greater dispersion and again nearly degenerate behavior at high-symmetry points.  These effects can be seen through comparison of phonon densities of states of Figure~1 and Figure~2, as well as through comparison of the absorption.  Figure~2 shows that the primary transparency band we calculate for GaP, just below 700 cm$^{-1}$, is observed.  Although this trough in the absorption appears weak next to the sharp IR-active phonon feature, it constitutes a local variation in the absorption by two orders of magnitude.  For both materials, spectral structure is also evident in the temperature-dependent far-infrared absorption below the reststrahlen, where critical points can be seen in the theoretical plots.

\indent  In short, we have obtained quite realistic results for the infrared spectra and their temperature dependence from first principles with an anharmonic two-phonon theory.  The behavior above twice the IR-active phonon cannot be modeled without higher-order multiphonon processes.




\begin{theacknowledgments}
We thank Michael E. Thomas, David W. Blodgett, Daniel V. Hahn and Simon G. Kaplan for the mid-IR GaP data and its analysis.
\end{theacknowledgments}




\begin{thebibliography}{8}
\expandafter\ifx\csname natexlab\endcsname\relax\def\natexlab#1{#1}\fi
\providecommand{\enquote}[1]{``#1''}
\expandafter\ifx\csname url\endcsname\relax
  \def\url#1{\texttt{#1}}\fi
\expandafter\ifx\csname urlprefix\endcsname\relax\def\urlprefix{URL }\fi

\bibitem[1]{genspec} G. Onida, L. Reining, and A. Rubio, Rev. Mod. Phys. {\bf 74}, 601 (2002).
\bibitem[2]{phonrev}S. Baroni, S. de Gironcoli, A. Dal Corso, and P. Giannozzi, Rev. Mod. Phys. {\bf{73}}, 515 (2001).
\bibitem[3]{Born}M. Born, Rep. Prog. Phys. {\bf{9}}, 294 (1943).
\bibitem[4]{Giannozzi}P. Giannozzi, S. de Gironcoli, P. Pavone, and S. Baroni, Phys. Rev. B 43, 7231 (1991).
\bibitem[5]{Thomas1}M.E. Thomas, D. Blodgett, D. Hahn, and S. Kaplan, SPIE Proceedings {\bf 5078}, Windows and Domes Technologies VIII, April 22-23 (2003).
\bibitem[6] {PalikAs} E.D. Palik, {\it Handbook of Optical Constants of Solids}, edited by E.D. Palik, (Academic, London, 1998).
\bibitem[7] {PalikP} A. Borghesi and G. Guizzetti, {\it Handbook of Optical Constants of Solids}, edited by E.D. Palik, (Academic, London, 1998).
\bibitem[8]{Stolen}R.H. Stolen, Phys. Rev. B {\bf{11}}, 767 (1975).
\bibitem[9]{Vin}V.S. Vinogradov, Sov. Phys.--Solid State {\bf{4}}, 519 (1962).
\bibitem[10]{Mar2}R.F. Wallis and A.A. Maradudin, Phys. Rev. {\bf{125}}, 1277 (1962).
\bibitem[11]{Cowl2}R.A. Cowley, Adv. Phys. {\bf{12}}, 421 (1963).
\bibitem[12]{Deb1}A. Debernardi, S. Baroni, and E. Molinari, Phys. Rev. Lett. {\bf{75}}, 1819 (1995).
\bibitem[13]{Lang}G. Lang, K. Karch, M. Schmitt, P. Pavone, A.P. Mayer, R.K. Wehner, and D. Strauch, Phys. Rev. B {\bf{59}}, 6182 (1999).
\bibitem[14]{Deb2}A. Debernardi, Solid State Commun. {\bf{113}}, 1 (2000).
\bibitem[15]{GaP}F. Widulle, T. Ruf, A. G\"{o}bel, E. Sch\"{o}nherr, and M. Cardona, Phys. Rev. Lett. {\bf {82}}, 5281 (1999).
\bibitem[16]{GaN}K.J. Yee, K.G. Lee, E. Oh, D.S. Kim, and Y.S. Lim, Phys. Rev. Lett. {\bf{88}}, 105501 (2002).
\bibitem[17]{AlAs}M. Canonico, C. Poweleit, J. Men\'endez, A. Debernardi, S.R. Johnson, and Y.H. Zhang, Phys. Rev. Lett. {\bf{88}}, 215502 (2002).
\bibitem[18]{Van}R.D. King-Smith and D. Vanderbilt, Phys. Rev. B {\bf{47}}, 1651 (1993).
\bibitem[19]{Resta}R. Resta, Rev. Mod. Phys. {\bf{66}}, 899 (1994).
\bibitem[20]{Gonze}X. Gonze and C. Lee, Phys. Rev. B {\bf{55}}, 10 355 (1997).
\bibitem[21]{Kubo} R. Kubo, J. Phys. Soc. Japan {\bf{12}}, 570 (1957).
\bibitem[22]{Cowl1}R.A. Cowley, Proc. Phys. Soc. {\bf{84}}, 281 (1964).
\bibitem[23]{Maradudin}A.A. Maradudin, E.W. Montroll, G.H. Weiss, and I.P. Ipatova, in {\it{Solid State Physics:  Advances in Research and Applications}}, edited by H.E. Ehrenreich, F. Seitz, and D. Turnbull (Academic, New York, 1971),  Suppl. 3, Chap. 6.
\bibitem[24]{Hove}L. Van Hove, N.M. Hugenholtz, and L.P. Howland, in {\it{Quantum Theory of Many-Particle Systems}}, (W.A. Benjamin, New York, 1961).

\bibitem[25]{Mar3}A.A. Maradudin and A.E. Fein, Phys Rev. {\bf{128}}, 2589 (1962).
\bibitem[26]{Kokkaddee}J. Kokkeddee, Physica {\bf{28}}, 374 (1962).
\bibitem[27]{handbuch}H. Bilz, D. Strauch, and R.K. Wehner, in {\it{Handbuch der Physik, Bd. XXV/2d:  Licht und Materie}}, edited by S. Fl\"{u}gge (Springer, Berlin, 1984).
\bibitem[28]{Deb3}A. Debernardi, M. Alouani, and H. Dreysse, Phys. Rev. B {\bf{63}}, 064305 (2001).
\bibitem[29]{Strauch}G. Deinzer, G. Birner, and D. Strauch, Phys. Rev. B {\bf{67}}, 144304 (2003).
\bibitem[30]{Cowl3}G. Dolling and R.A. Cowley, Proc. Phys. Soc. {\bf{88}}, 463 (1966).
\bibitem[31]{Lax}M. Lax and E. Burstein, Phys Rev. {\bf 97}, 39 (1955).
\bibitem[32]{Strauchtp}G. Deinzer and D. Strauch, Phys. Rev. B {\bf{69}}, 045205 (2004).
\bibitem[33]{Thomas}M.E. Thomas, R.I. Joseph, and W.J. Tropf, Appl. Opt. {\bf 27}, 239 (1988).
\bibitem[34]{YandC}M.~T. Yin and M.~L. Cohen, Phys. Rev. B {\bf 26}, 3259 (1982).
\bibitem[35]{us}H.M. Lawler, E.K. Chang, and E.L. Shirley, unpublished.
\bibitem[36]{HK}P. Hohenberg and W. Kohn, Phys. Rev. {\bf 136}, B864 (1964); W. Kohn and L.~J. Sham, Phys. Rev.
{\bf 140}, A1133 (1965); J. Perdew and A. Zunger, Phys. Rev. B {\bf 23}, 5048 (1981).
\bibitem[37]{Hamman}D.~R. Hamann, M. Schl\"{u}ter, and C. Chiang, Phys. Rev. Lett. {\bf 43}, 1494 (1979).
\bibitem[38]{Vanderbilt}D. Vanderbilt, Phys. Rev. B {\bf 32}, 8412 (1985).
\bibitem[39]{KandB}L. Kleinman and D.~M. Bylander, Phys. Rev. Lett. {\bf 48}, 1425 (1982).
\end{thebibliography}

\IfFileExists{\jobname.bbl}{}
 {\typeout{}
  \typeout{******************************************}
  \typeout{** Please run "bibtex \jobname" to optain}
  \typeout{** the bibliography and then re-run LaTeX}
  \typeout{** twice to fix the references!}
  \typeout{******************************************}
  \typeout{}
 }

\end{document}